\documentstyle[12pt]{article}
\title{ Topologimeter and the problem of physical interpretation of
topology lattice.}

\author{ A. A. Grib, R. R. Zapatrin\\
A. A. Friedmann Laboratory For Theoretical Physics\\
Sadovaya, 21, \\
191023, St-Petersburg, Russia.}

\date{}
\newtheorem{Lemma}{Lemma}
\begin{document}
\maketitle
\begin{abstract}
The collection of all topologies on the set of three points is
studied treating the topology as quantum-like observable. It turns
out to be possible under the assumption of the asymmetry between
the spaces of bra- and ket-vectors. The analogies between the
introduced topologimeter and the Stern-Gerlach experiments are
outlined.
\end{abstract}

\section*{ Introduction. }

An important problem in modern theoretical physics is the problem of
quantum topology. What is quantum topology? Can one have something
like a wave function for topologies? Can one speak about probability
of this or that topology,
and is it possible to speak about probability calculus for topologies?
Are "quantum jumps" between different topologies possible?
It is well known that in classical physics one can use classical
probability measure, but for molecules, atoms, elementary particles
and may be gravity one has probability amplitude description
in terms of wave functions. This is connected with the fact that
the lattice of properties of quantum system is non distributive
(non-Boolean) (Birkhoff and von Neumann, 1936),  and one has the
formalism of Hilbert space, noncommutative operators, wave function
and Heisenberg uncertainty relations.

One can speak about quantum topology if one deals with Planckean
scales for space-time when gravity must be quantised. There is
other interesting possibility to speak about quantum topology
when according to Leinaas and Myrheim (1991) one gives the
interpretation of Pauli principle in terms of nontrivial topology
for configuration space of many particle system, so that in EPR
experiment, when one goes from symmetrised two particle to the
product after measuring local observables one has change of
topologies.  Here we shall investigate the most simple case of
discrete topologies.  From (Zapatrin, 1993; Sorkin, 1991) one can
see that our analysis can give some insight for continuous case as
well.

\section{Topologimeter}

In this paper we shall study the properties of the topology lattice
for three points, continuing our investigation in (Grib and
Zapatrin, 1992).  The striking feature of this lattice and of
lattices of topologies for the number of points $n$ greater or
equal to 3, is that they are nondistributive.  This makes it
impossible to have classical measure on these lattices and leads to
some resemblance of them to quantum systems, but without Planck's
constant. As in (Grib and Zapatrin, 1992), consider the work of a
hypothetical 'topologimeter' --- an apparatus which can ask and
obtain answers on the question: "what is the topology of the
three-point set?"  We come to the conclusion that due to
nondistributivity of the lattice one obtains noncausal "quantum
jumps" of topologies similar to situation in Stern-Gerlach
experiment when complementary observables are measured.  Really,
consider the triple $a$, $(ac)$, $(ab)$ of atomic topologies
(the notations are the same as in Grib and
Zapatrin, 1992):
\begin{equation}
a\land ((ac)\lor (ab))\quad\hbox{ is not equal
to }\quad (a\land (ac))\lor (a\land (ab))
\label{0nd}\end{equation}
Here $(ac)\lor (ab)$ for our topologimeter is also some topology
defined following Finkelstein (1963) for the case of quantum logic
in the sense that if $(ac)$ is "true" then $((ac)\lor(ab))$ is
true, if $(ab)$ is true then $(ac)\lor (ab)$ true.

The Boolean-minded observer comes to the following result using
the topologimeter.  Imagine he sees $a$ is true. Then, since
$a=a\land (ac)\lor (ab))$ he will say according to his Boolean
distributive logic interpreting $\land$ as "and", $\lor$ as "or"
that $(ac)\lor (ab)$ is true. Then he must say that either $(ac)$
or $(ab)$ is true. However, the atomic topologies $a$, $(ac)$,
$(ab)$ are incompatible since from Hasse diagram we see $a\land
(ac)=(ac)\land (ab)=a\land (ab)=0$.  The escape from this
contradiction is to assume the 'non-causal jump of the topology in
different moments of time'. The time here plays the crucial role.
If $a$ is true for $t=t_0$, then the Boolean-minded observer will
say that at some other moment $t=t_1$ the topology can non-causally
"jump" into some $(ac)\lor (ab)$.

So the situation here resembles that for measuring
noncommuting spin operators in the Stern-Gerlach experiment.
To make this analysis more close to quantum mechanics
and to obtain some "quantum topology" one can try to find
something like matrix representation of our lattice of
topologies to compare it with usual Pauli matrices for
spin. In this paper we show that such representation
can be constructed. But differently from quantum mechanics
one can not have formulation in terms of one Hilbert space
and the wave function there. One must have two spaces
and our matrices are operators from one space to the
other. There are both commuting and non commuting matrices.
All this shows that here we have the example of a new
system, different from classical as well as from
quantum mechanical cases.

Our matrix  representation  shows  that there is the direct
correspondence between noncommutativity of operators  and
nondistributivity  of  the lattice.

For nondistributive triples $ a,  (ac), (ab); b, (bc), (ab); c, (bc),
(ac)$ one has representation  in  terms  of  noncommuting  "bra
operators" (matrices): $a$  doesn't  commute  with  $(ac)$ and
$(ab)$,  $b$ with $(bc)$ and $(ab)$, same with $c$.

There are distributive triples like $a ,  b ,  c$ , or
$(ab),(ac),(bc)$.  That is  why  comparing the  work of our
"topologimeter"  with Stern--Gerlach experiment one can predict the
following.  If the topology  is fixed as "$a$"("$a$" is true) then
the prediction that at the next moment if the question will be
asked:"is \lq $(ac)$\rq true?" --- the answer can be "yes" --- the
topology will jump from "$a$" to "$(ac)$". The same is true if
the question will be about $(ab)$.  But if the question will
be:"if at $t=t_0$ "$a$"  is true,  is "$b$" true at the next
moment?" --- the answer will be "no". The same prediction will be
about "$c$". The situation here resembles measuring  of  two
noncommuting spin projections for the quantum particle with spin 1.
For spin 1 case there are three commuting  projectors $S_x=
1,S_x= 0,S_x=-1$ which do not commute with $S_y= 1,S_y=0,S_y=-1$.

\begin{figure}[htb]
\unitlength=2.12mm

\linethickness{0.4pt}
\begin{picture}(55,65)

\put(30,5){\makebox(0,0)[cc]{$0$}}

\put(30,6){\line(1,1){7}}
\put(31,6){\line(2,1){15}}
\put(32,6){\line(3,1){25}}
\put(30,6){\line(-1,1){7}}
\put(29,6){\line(-2,1){15}}
\put(28,6){\line(-3,1){25}}

\put(3,15.05){\makebox(0,0)[cc]{$a$}}
\put(13,15.05){\makebox(0,0)[cc]{$(ac)$}}
\put(23,15.05){\makebox(0,0)[cc]{$c$}}
\put(37,15.05){\makebox(0,0)[cc]{$(bc)$}}
\put(47,15.05){\makebox(0,0)[cc]{$b$}}
\put(57,15.05){\makebox(0,0)[cc]{$(ab)$}}

\put(3,17){\line(0,1){7}}
\put(3,17){\vector(-1,1){2}}
\put(2,18){\vector(-1,1){2}}
\put(1,19){\vector(-1,1){2}}
\put(13,17){\line(0,1){7}}
\put(11,17){\line(-1,1){7}}
\put(23,17){\line(0,1){7}}
\put(21,17){\line(-1,1){7}}
\put(37,17){\line(0,1){7}}
\put(34,17){\line(-1,1){8}}
\put(47,17){\line(0,1){7}}
\put(45,17){\line(-1,1){7}}
\put(57,17){\line(0,1){7}}
\put(55,17){\line(-1,1){7}}
\put(61,20){\vector(-1,1){2}}
\put(60,21){\vector(-1,1){2}}
\put(59,22){\vector(-1,1){2}}

\put(4,26){\makebox(0,0)[cc]{$a(ac)$}}
\put(14,26){\makebox(0,0)[cc]{$c(ac)$}}
\put(24,26){\makebox(0,0)[cc]{$c(bc)$}}
\put(36,26){\makebox(0,0)[cc]{$b(bc)$}}
\put(46,26){\makebox(0,0)[cc]{$b(ab)$}}
\put(56,26){\makebox(0,0)[cc]{$a(ab)$}}

\put(3,28){\line(0,1){7}}
\put(3,28){\vector(-1,1){2}}
\put(2,29){\vector(-1,1){2}}
\put(13,28){\line(0,1){7}}
\put(11,28){\line(-1,1){7}}
\put(23,28){\line(0,1){7}}
\put(21,28){\line(-1,1){7}}
\put(37,28){\line(0,1){7}}
\put(34,28){\line(-1,1){7}}
\put(47,28){\line(0,1){7}}
\put(45,28){\line(-1,1){7}}
\put(57,28){\line(0,1){7}}
\put(55,28){\line(-1,1){7}}
\put(62,31){\vector(-1,1){2}}
\put(60,33){\vector(-1,1){2}}

\put(27,30){\makebox(0,0)[cc]{$a(bc)$}}
\put(38,30){\makebox(0,0)[cc]{$b(ac)$}}
\put(48,30){\makebox(0,0)[cc]{$c(ab)$}}
\put(3,17){\line(2,1){24}}
\put(13,17){\line(2,1){24}}
\put(23,17){\line(2,1){24}}
\put(34,17){\line(-1,2){6}}
\put(46,17){\line(-1,2){6}}
\put(56,17){\line(-1,2){6}}
\put(5,45){\line(3,-2){20}}
\put(15,45){\line(3,-2){20}}
\put(27,45){\line(3,-2){20}}
\put(34,45){\line(-1,-2){7}}
\put(46,45){\line(-1,-2){7}}
\put(56,45){\line(-1,-2){7}}

\put(4,37){\makebox(0,0)[cc]{$ac(ac)$}}
\put(14,37){\makebox(0,0)[cc]{$c(ac)(bc)$}}
\put(24,37){\makebox(0,0)[cc]{$bc(bc)$}}
\put(38,37){\makebox(0,0)[cc]{$b(ab)(bc)$}}
\put(48,37){\makebox(0,0)[cc]{$ab(ab)$}}
\put(58,37){\makebox(0,0)[cc]{$a(ab)(ac)$}}

\put(3,38){\line(0,1){7}}
\put(3,38){\vector(-1,1){4}}
\put(13,38){\line(0,1){7}}
\put(11,38){\line(-1,1){7}}
\put(23,38){\line(0,1){7}}
\put(21,38){\line(-1,1){7}}
\put(37,38){\line(0,1){7}}
\put(34,38){\line(-1,1){7}}
\put(47,38){\line(0,1){7}}
\put(45,38){\line(-1,1){7}}
\put(57,38){\line(0,1){7}}
\put(55,38){\line(-1,1){7}}
\put(62,41){\vector(-1,1){4}}

\put(4,47){\makebox(0,0)[cc]{$ac(ac)(bc)$}}
\put(14,47){\makebox(0,0)[cc]{$bc(ac)(bc)$}}
\put(24,47){\makebox(0,0)[cc]{$bc(ab)(bc)$}}
\put(38,47){\makebox(0,0)[cc]{$ab(ab)(bc)$}}
\put(48,47){\makebox(0,0)[cc]{$ab(ab)(ac)$}}
\put(58,47){\makebox(0,0)[cc]{$ac(ab)(ac)$}}

\put(30,57){\line(1,-1){7}}
\put(31,57){\line(2,-1){15}}
\put(32,57){\line(3,-1){25}}
\put(30,57){\line(-1,-1){7}}
\put(29,57){\line(-2,-1){15}}
\put(28,57){\line(-3,-1){25}}

\put(30,58){\makebox(0,0)[cc]{$I=abc$}}
\end{picture}
\caption{ The lattice $\tau (3)$.}
\label{t3}
\end{figure}
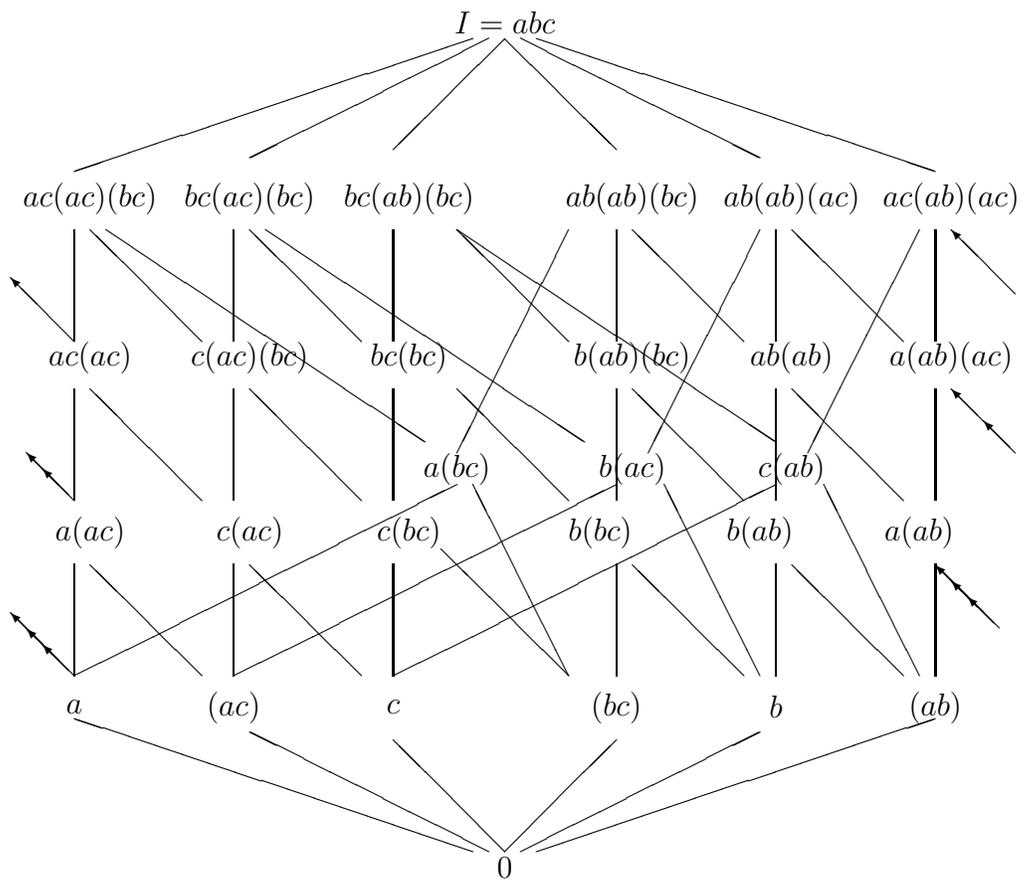

Nevertheless our lattice differs from quantum logical
lattice of spin 1 particle as we have shown it earlier (Grib and
Zapatrin, 1992).  It is easy to  see  from Fig.\ref{t3} that
not only $(b),(c)$  commute with $(a)$ but also
the atomic $(bc)$, not commuting with $(b)$ and $(c)$.

There is no natural orthogonality in the lattice so if one imposes some
orthogonality  in the 6-dimensional  space ${\cal H}_V$ by defining
a scalar product and treating mutually commuting idempotents as
orthogonal projectors one  obtains the contradictory system of
equations.  So, orthogonal subspaces may not correspond to
projectors on any topology represented in the  lattice
Fig.\ref{t3}.  This can be treated as some superpositions of
topologies which are not observable by our topologimeter.

The most interesting aspect which makes  this  system  different  from
usual quantum microparticle is that it is "classical" ---
"macroscopic".  There is no need for macroscopical apparatus to
measure  complementary properties of  some  microparticle.  This
shows  that  the reason for "jumps" corresponding to wave packet
collapse in quantum mechanics  is not connected  with any
intervention of macroscopical apparatus but is due to
interpretation of nondistributive lattice by some consciousness
with Boolean logics. Asking question, using Boolean logics, consciousness
interprete non-Boolean structure in such a way that it uses  time
in order  to  charge "yes-no" values for topologies as it is also done
in quantum logic.

But contrary to usual Stern-Gerlach experiment for spin 1  case  there
is no  need  for two different complementary topologimeters for
measuring noncommuting operators.  It is enough to have  one  topologimeter
but one  must look for two different moments of time in order to check
values of complementary topologies.

The absence of different classical measuring apparatuses for this
case leads to  absence of necessity to have different von
Neumann's measuring Hamiltonian (von Neumann, 1955) for
complementary observables.  There is no "interaction" between
Boolean consciousness  and non-Boolean system in this case.  This
makes the investigation of non-Boolean lattice of topologies
comparing it with usual quantum logical system important for our
understanding of the deep problem of the role of consciousness in
measurment theory in quantum physics either.

\section{REPRESENTATION OF THE TOPOLOGY LATTICE $\tau(3)$
BY OPERATORS}

The draft scheme of the construction  is  the following.  We intend
to represent the elements of the topology lattice $L=\tau(3)$ by
operators in a linear space. Two 6-dimensional spaces ${\cal
H}_{V}$  and ${\cal H}_{\Lambda }$ called {\bf bra-space} and {\bf
ket-space}, respectively are  considered.  The basis of ${\cal
H}_{V}$ is labelled by atoms of $L$, and  the  basis  of ${\cal
H}_{\Lambda }$  is labelled by coatoms of {\it L}. Then each
element of $L$ has  the  twofold representation: as subspace of
${\cal H}_{V} ($called {\it bra- representation})  or as that of
${\cal H}_{\Lambda } (${\it ket-representation}). The meet
operation  in $L$  is easily described in  terms  of  the  bra-
representation  as  the set-theoretic intersections of appropriate
subspaces. The joins in $L$  are  associated  with  the  set
intersections  in   the   ket- representation. The object of the
mathematical  treatise  exposed below is the construction of joins
in terms of ${\cal H}_{V}$  and  meets  in terms of ${\cal
H}_{\Lambda }$.

\subsection{Bra- and ket- representations}

Let $L=\tau(3)$  be  the  lattice  of  all  topologies  on  a  set
$X=\{a,b,c\}$ of 3 points. The atoms  of $L$  are  the  proper  weakest
topologies on $X$, each of which associated  with  the  only  proper
(that is, not equal to $\emptyset$ or $X$) open subset of
$X=\{a,b,c\}$.  Denote by ${\cal H}_{V}$ the 6-dimensional linear
space with the basis labelled by these subsets:
\[ \cal H =
{\rm span}\{{\bf e}_a,{\bf e}_{(ab)},{\bf e}_b,{\bf e}_{(bc)}, {\bf
e}_c,{\bf e}_{(ac)}\}
\]
The  lattice $L=\tau(3)$  is
CAC (complete  atomistic   coatomistic)
lattice (Larson and Andima, 1975), that is why each topology
$\tau\in L$ is the join  of atomic topologies which are weaker than
$\tau$.  So, $\tau$ can  be  unambiguously associated with the
subspace $V_{\tau}\subseteq {\cal H}_{V}$ spanned  on  the
appropriate basis vectors:

\begin{equation} V_\tau = {\rm
span}\{{\bf e}_A\mid\quad A\in\tau\}\label{vtau}
\end{equation}

\noindent In other words, for any subset $A\subseteq X$, ${\bf
e}_{A}\in V_{\tau}$ if and only if the set $A$ is open
in the topology $\tau$.

The ket-representation of $L$ is built likewise.  The  coatoms
of $L$  are  associated  with  the  maximal  proper  topologies  on
$X=\{a,b,c\}$ each of which is associated  with  an  ordered
pair  of elements of $X$(Zapatrin, 1993a), denoted
by $a\rightarrow b$, $c\rightarrow a$,  etc.  For example:

$$
a\rightarrow b\hbox{ is }\{\emptyset ,\{b\},\{c\},\{b,c\},\{a,b\},X\}
$$

$$ c\rightarrow a\hbox{ is }\{\emptyset ,\{a\},\{b\},\{a,b\},\{a,c\},X\}
$$

\noindent and so
on. Now introduce the 6-dimensional ket-space ${\cal H}_{\Lambda }$
with  the basis labelled by ordered pairs of $X=\{a,b,c\}$:
\[ {\cal H}_{\Lambda }=\hbox{span}\{{\bf f}_{ab},{\bf f}_{ba},{\bf
f}_{ac},{\bf f}_{ca},{\bf f}_{bc},{\bf f}_{cb}\} \]

Each topology $\tau\in L$ can now  be  represented  as  the
subspace $\Lambda _{\tau}\subseteq {\cal H}_{\Lambda }$
spanned on the basis vectors ${\bf f}_{xy}$ , $x,y\in X$
satisfying  the following condition:

\begin{equation}
{\bf f}_{xy}\in\Lambda_\tau\quad\Leftrightarrow\quad (\forall A\in\tau\quad
x\in A\Rightarrow y\in A)\label{fxy}
\end{equation}

\subsection{Examples}

Six examples of topologies are presented on the Table \ref{ttop}.

\begin{table}[htb]
{\tenrm
\begin{tabular}{l||l||l}\hline
Topology $\tau$\qquad & Bra-representation & Ket-representation \\
\hline
$a=\{\emptyset ,\{a\},X\}$&$V_{a}={\rm span}\{{\bf e}_{a}\}$
&$\Lambda _{a}={\rm span}\{{\bf f}_{ba}{\bf f}_{ca}{\bf f}_{bc}{\bf
f}_{cb}\}$\\
(atomic topology)  &  $\dim{\it V}_{a}=1$             & $\dim
\Lambda _{a}=4$\\ \hline
$=\{\emptyset,\{a,b\},X\}$ & $V_{\tau}=\hbox{span}\{{\bf
e}_{(ab)}\}$ &
 $\Lambda _{\tau}={\rm span}\{{\bf f}_{ba}{\bf
f}_{ca}{\bf f}_{ab}{\bf f}_{cb}\}$\\
(atomic topology)  &  $\dim V_{(ab)}=1$          & $\dim
\Lambda _{(ab)}=4$\\ \hline
$ a\vee b = ab(ab) =$ &
$V_{\tau}={\rm span}\{{\bf e}_{a},{\bf e}_{b},{\bf e}_{(ab)}\}$ &
$\Lambda _{\tau}={\rm span}\{{\bf f}_{ca}{\bf f}_{cb}\}$\\
$= \{\emptyset ,a,b,\{a,b\},X\}$  & $\dim V_{\tau}=3\neq
\dim V_{a}+\dim V_{b}$  &     $\dim \Lambda
_{\tau}=2$\\ \hline
$a(bc)=\{\emptyset ,a,\{b,c\},X\}$&
 $V_{\tau}={\rm span}\{{\bf
e}_{a},{\bf e}_{(bc)}\}$ & $\Lambda _{\tau}={\rm span}\{{\bf
f}_{bc}{\bf f}_{cb}\}$\\
&   $\dim V_\tau =2\qquad$ & $\dim \Lambda _{\tau}=2$\\
\hline
$(a\rightarrow b)=\{\emptyset ,b,c,\{b,c\},$ &
$V_{\tau}={\rm span}\{{\bf e}_{b},{\bf e}_{c},\qquad$ & $\Lambda
_{\tau}={\rm span}\{{\bf f}_{ab}\}$\\
$\{a,b\},X\}\qquad$ & ${\bf e}_{(ab)},{\bf e}_{(bc)}\}\qquad$
&         \\
(coatomic topology)&   $\dim V_{\tau}=4$            &
$\dim \Lambda _{\tau}=4$\\ \hline
\end{tabular}
\label{ttop}
\caption{Some examples of topologies}
}
\end{table}

\subsection{Transition between the representations}

Let $\tau\in L$ is represented by the bra-subspace
$V_{\tau}\subseteq {\cal H}_{V}$.  Denote  by $P_{\tau}$ the
projector onto $V_{\tau}$. Let us  construct  the  algorithm  which
builds $\Lambda _{\tau}$ by given $V_{\tau}$,  and  vice  versa.
To  proceed  it,  first introduce the {\bf sandwich operator}
$S:{\cal H}_{V}\rightarrow {\cal H}_{\Lambda }$ in matrix form as:
\[ S_{\lambda v}=\cases{0 &
,$v\le\lambda$ (i.e. the atom $v$ is below the coatom $\lambda$)\cr
1 &, otherwise}
\]

Its transposed $S^{T}$ will be the operator $S^{T}:{\cal
H}_{\Lambda }\rightarrow {\cal H}_{V}$. Consider the product
$SP_{\tau}:{\cal H}_{V}\rightarrow {\cal H}_{\Lambda }$. In
the space ${\cal H}_{\Lambda }$, denote  by $\Pi _{\tau}$ the
projector onto the subspace $\Lambda _{\tau}$. In
particular,  denote  the projectors onto the basis vectors
${\bf e}_{xy}, x,y\in \{a,b,c\}$ by $\pi _{xy}$.

\begin{Lemma}\label{R.1} Let $\tau\in L$ be a topology
on $X=\{a,b,c\}$, $V_{\tau}$ be its bra-representation, and
$P_{\tau}$ be the projector in ${\cal H}_{V}$ associated
with $V_{\tau}$. Then the projector $\Pi _{\tau}$ in
${\cal H}_{\Lambda }$ associated with $\Lambda _{\tau}$
is:

\begin{equation} \Pi_\tau =
\Sigma\{\pi_{xy}\mid\quad\pi_{xy}SP_\tau = 0\}\label{pitau}
\end{equation}
\end{Lemma}

\noindent {\it Proof.} $\pi_{xy}$ is included in the sum
({\ref{pitau}}) if and only if $S(xy,v)=0$ for any
atomic topology $v$ which is weaker then $\tau$. Due to
(R.3) that means that $\pi _{xy}\le \Pi
_{\tau}$ iff

$$
\forall v v\le \tau \Leftrightarrow  v\le \lambda
$$

\noindent Since the lattice $L$ is coatomistic, $\tau=\wedge
\{\lambda \mid \tau\le \lambda \}$, thus (\ref{fxy}) implies
(\ref{pitau}).

The following "transposed" Lemma is proved  likewise.  Denote
by $p_{A}$ the projector onto the vector ${\bf e}_{A}$ in ${\cal H}_{V}$.

\begin{Lemma}\label{R.2}  The   transition   from   the   ket-
to bra- representation is described as follows:
\[
P_{\tau}=\sum \{p_{A}\mid  p_{A}S^{T}\Pi _{\tau}=0\}\qquad (\hbox{R.}5)
\]
\end{Lemma}

\subsection{Lattice joins in bra--representation}

Let $\sigma ,\tau\in L$, and let their bra--representations  be
$V_{\sigma },V_{\tau}\subseteq {\cal H}_{V}$, associated with the
projectors $P_{\sigma },P_{\tau}$, respectively. To  build  the
projector $P_{\sigma \vee \tau}$, perform consecutively  the
transition  procedures described  in  the   Lemmas \ref{R.1 , R.2}.
First   form   the   ket--representation $\Lambda _{\sigma \vee
\tau}$ associated with the projector (\ref{pitau}):

\begin{equation}
\Pi _{\sigma \vee \tau}=\sum \{\pi _{xy}\mid  \pi
_{xy}S(P_{\sigma }+P_{\tau})=0\}
\label{R.6}
\end{equation}

\noindent and then go  backwards  to ${\cal H}_{V}$. (R.6)  is
really  the  projector associated with the join since it is the
meet of all upper  bounds for both $\sigma $ and $\tau$ (Zapatrin,
1994).

\subsection{An example}

Let us explicitly build the projector associated with
the join of two atomic topologies $a$ and $b$.

$$
{\bf e}_a=\left[ \begin{array}{cccccc}1\\0\\0\\0\\0\\0\\
\end{array}\right]
\quad\Rightarrow\quad P_a=\left[\begin{array}{cccccc}
 1 & 0 & 0 & 0 & 0 & 0\\
 0 & 0 & 0 & 0 & 0 & 0\\
 0 & 0 & 0 & 0 & 0 & 0\\
 0 & 0 & 0 & 0 & 0 & 0\\
 0 & 0 & 0 & 0 & 0 & 0\\
 0 & 0 & 0 & 0 & 0 & 0
\end{array}\right]
$$

$$
{\bf e}_b=\left[ \begin{array}{ll}0\\0\\1\\0\\0\\0\\
\end{array}\right]
\quad\Rightarrow\quad P_b=\left[\begin{array}{cccccc}
0 & 0 & 0 & 0 & 0 & 0\\
0 & 0 & 0 & 0 & 0 & 0\\
0 & 0 & 1 & 0 & 0 & 0\\
0 & 0 & 0 & 0 & 0 & 0\\
0 & 0 & 0 & 0 & 0 & 0\\
0 & 0 & 0 & 0 & 0 & 0\end{array}\right]
$$

Set  up  the   following   order   of   basis   vectors   in ${\cal
H}_{V}$:  ${\bf e}_{a}$,${\bf e}_{ab},{\bf e}_{b},{\bf e}_{bc},{\bf
e}_{c},{\bf e}_{ac}$, and of those in ${\cal H}_{\Lambda }$:  ${\bf
f}_{ca},{\bf f}_{ba},{\bf f}_{bc},{\bf f}_{ac},{\bf f}_{ab},{\bf
f}_{cb}$, then the matrix of the sandwich operator is:
\begin{equation}
S=\left[\begin{array}{cccccc}
1 & 1 & 0 & 0 & 0 & 0\\
0 & 1 & 1 & 0 & 0 & 0\\
0 & 0 & 1 & 1 & 0 & 0\\
0 & 0 & 0 & 1 & 1 & 0\\
0 & 0 & 0 & 0 & 1 & 1\\
1 & 0 & 0 & 0 & 0 & 1\end{array}\right]
\label{1s}\end{equation} Then \[
P_a+P_b=\left[\begin{array}{cccccc}
1 & 0 & 0 & 0 & 0 & 0\\
0 & 0 & 0 & 0 & 0 & 0\\
0 & 0 & 1 & 0 & 0 & 0\\
0 & 0 & 0 & 0 & 0 & 0\\
0 & 0 & 0 & 0 & 0 & 0\\
0 & 0 & 0 & 0 & 0 & 0\end{array}\right] \qquad\hbox{, hence }\quad
S(P_a+P_b) = \left[\begin{array}{cccccc}
1 & 0 & 0 & 0 & 0 & 0\\
0 & 0 & 1 & 0 & 0 & 0\\
0 & 0 & 1 & 0 & 0 & 0\\
0 & 0 & 0 & 0 & 0 & 0\\
0 & 0 & 0 & 0 & 0 & 0\\
1 & 0 & 0 & 0 & 0 & 0\end{array}\right]
\] Therefore the only $\pi_{xy}$ satisfying \ref{R.6}
are  the  projectors  onto the following vectors \[{\bf
f}_{bc}=(0,0,1,0,0,0)\]\[{\bf f}_{ca}=(1,0,0,0,0,0)\]
hence
\[ \Pi_{a\lor b}=\pi_{bc}+\pi_{ca} =
\left[\begin{array}{cccccc}
1 & 0 & 0 & 0 & 0 & 0\\
0 & 0 & 0 & 0 & 0 & 0\\
0 & 0 & 1 & 0 & 0 & 0\\
0 & 0 & 0 & 0 & 0 & 0\\
0 & 0 & 0 & 0 & 0 & 0\\
0 & 0 & 0 & 0 & 0 & 0\end{array}\right] \] Then, it follows from
the Lemma \ref{R.2} that \[ S^T\Pi_{a\lor b} =
\left[\begin{array}{cccccc}
1 & 0 & 0 & 0 & 0 & 1\\
1 & 1 & 0 & 0 & 0 & 0\\
0 & 1 & 1 & 0 & 0 & 0\\
0 & 0 & 1 & 1 & 0 & 0\\
0 & 0 & 0 & 1 & 1 & 0\\
0 & 0 & 0 & 0 & 1 & 1\end{array}\right]
\circ
\left[\begin{array}{cccccc}
1 & 0 & 0 & 0 & 0 & 0\\
0 & 0 & 0 & 0 & 0 & 0\\
0 & 0 & 1 & 0 & 0 & 0\\
0 & 0 & 0 & 0 & 0 & 0\\
0 & 0 & 0 & 0 & 0 & 0\\
0 & 0 & 0 & 0 & 0 & 0\end{array}\right] =
\left[\begin{array}{cccccc}
1 & 0 & 0 & 0 & 0 & 0\\
1 & 0 & 0 & 0 & 0 & 0\\
0 & 0 & 1 & 0 & 0 & 0\\
0 & 0 & 1 & 0 & 0 & 0\\
0 & 0 & 0 & 0 & 0 & 0\\
0 & 0 & 0 & 0 & 0 & 0\end{array}\right]
$$

The $p_{A}$'s satisfying (\ref{R.6}) are the
projectors onto the vectors:

$$
{\bf e}_a=\left[\begin{array}{ll}1\\0\\0\\0\\0\\0\end{array}\right]\qquad
{\bf e}_{(ab)}=\left[\begin{array}{ll}0\\1\\0\\0\\0\\0\end{array}\right]\qquad
{\bf e}_b=\left[\begin{array}{ll}0\\0\\1\\0\\0\\0\end{array}\right]
\]Hence \[\Pi_{a\lor b}=\pi_a+\pi_{ab}+\pi_b
= \left[\begin{array}{cccccc}
1 & 0 & 0 & 0 & 0 & 0\\
0 & 1 & 0 & 0 & 0 & 0\\
0 & 0 & 1 & 0 & 0 & 0\\
0 & 0 & 0 & 0 & 0 & 0\\
0 & 0 & 0 & 0 & 0 & 0\\
0 & 0 & 0 & 0 & 0 & 0\end{array}\right] \neq \pi_a+\pi_b
\]

The other joins in ${\cal H}_{V}$ are built in the same way.

\section{Commutation relations}

At first  sight, the  proposed  representation  of  property
lattices seems  inconsistent  with  quantum  mechanical  intuition
since the operators associated with the atoms of property  lattice
all commute.  To reason about the  commutativity  of  observables
we  must somehow take into account both bra-  and
ket--representation  at once.

For instance, consider the pair of operators $P_a$ and $P_{(ab)}$.
They act from the bra-space to the ket-space, therefore it makes no
sense to speak about they commutation since they can not be
multiplied. To speak about commutation relations, we have to render
them to the same space. Note that we already have the operator
doing it, namely, the sandwich matrix $S$ (\ref{1s}), and the
operators $SP_a$ and $SP_{(ab)}$ will already act in the same space
having the form:
\[
SP_a = \left[\begin{array}{cccccc}
1 & 0 & 0 & 0 & 0 & 0\\
0 & 0 & 0 & 0 & 0 & 0\\
0 & 0 & 0 & 0 & 0 & 0\\
0 & 0 & 0 & 0 & 0 & 0\\
0 & 0 & 0 & 0 & 0 & 0\\
1 & 0 & 0 & 0 & 0 & 0\end{array}\right]
\qquad\mbox{ ; }\qquad
SP_{(ab)} = \left[\begin{array}{cccccc}
0 & 1 & 0 & 0 & 0 & 0\\
0 & 1 & 0 & 0 & 0 & 0\\
0 & 0 & 0 & 0 & 0 & 0\\
0 & 0 & 0 & 0 & 0 & 0\\
0 & 0 & 0 & 0 & 0 & 0\\
0 & 0 & 0 & 0 & 0 & 0\end{array}\right]
\]

\noindent It can be checked directly that both they are
idempotents: $(SP_{a})^2=SP_{a}$ and $(SP_{(ab)})^2=SP_{(ab)}$. We
could try to make them projectors (that is, self adjoint operators)
by introducing a scalar product in the bra-space in , say, usual
Euclidean way. Although, unlike the quantum mechanical situation,
they will not be self adjoint: $(SP_{a})^T\neq SP_{a}$,
$(SP_{(ab)})^T=SP_{(ab)}$.

We have to introduce commutation relation in such a way that they
could grasp the entire structure of the property lattice. When we
try to use the introduced above projectors $P_a, P_b$ etc., we
immediately see that they all commute which contradicts to the
violation of distributivity (\ref{0nd}). The idea we put forth is
the following. Since all the operators associated with the elements
of the lattice act from ${\cal H}_V$ to ${\cal H}_\Lambda$, we can
render them into one space, namely  ${\cal H}_V$, by multiplying
all of them by the matrix $S$ from the right side.
Then define the new product $\circ$
of operators in ${\cal H}_V$ \[ A\circ B:=ASBS \] and calculate all
commutators $[P_u, P_v]$ for all $u,v\in V$.

\begin{Lemma} For any $u,v\in V$
\begin{equation} [P_u,P_v] = S_{uv}(P_u - P_v) = \left\{
\begin{array}{ccl}
= & 0 & \mbox{if }S_{uv}=0\cr
\neq & 0 &\mbox{otherwise}
\end{array}\right.
\label{1nc}
\end{equation}
\end{Lemma}

\noindent {\it Proof\/} is obtained from direct checking by
multiplication of appropriate matrices.

The results of the calculation are shown on the Figure \ref{1com}
where the vertices associated with commuting projectors are linked
by lines of the diagram, and the pairs of vertices which are not
connected by a line, are associated with non-commuting projectors.
Now the correspondence between commutativity of projectors and
distributivity of the elements of the lattice is gathered. For
instance, the atomic topologies $a,b,c$ form the distributive
triple, and the appropriate projectors pairwise commute.

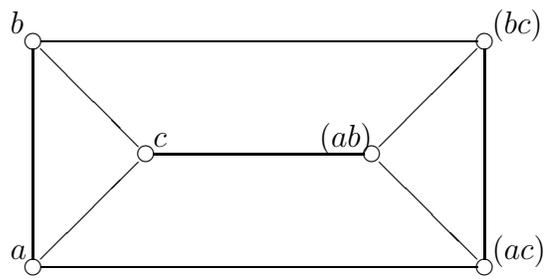
\begin{figure}[htb]
\unitlength 1mm
\begin{center}
\begin{picture}(80,50)
\put(0,0){\circle{2}}
\put(-3,1){\mbox{$a$}}
\put(0,30){\circle{2}}
\put(-3,31){\mbox{$b$}}
\put(15,15){\circle{2}}
\put(16,16){\mbox{$c$}}
\put(60,0){\circle{2}}
\put(61,31){\mbox{$(bc)$}}
\put(45,15){\circle{2}}
\put(38,16){\mbox{$(ab)$}}
\put(60,30){\circle{2}}
\put(61,1){\mbox{$(ac)$}}
\put(1,0){\line(1,0){58}}
\put(16,15){\line(1,0){28}}
\put(1,30){\line(1,0){58}}
\put(0,1){\line(0,1){28}}
\put(60,1){\line(0,1){28}}
\put(1,1){\line(1,1){13}}
\put(1,29){\line(1,-1){13}}
\put(46,16){\line(1,1){13}}
\put(46,14){\line(1,-1){13}}
\end{picture}
\end{center}
\caption{Commutativity of atomic projectors}
\label{2com}
\end{figure}

\section*{ Concluding remarks }

So, we see that the matrix representation of topology lattice
for three points is possible if one uses two spaces,
called "bra"-- and "ket"--spaces. Contrary to usual quantum
mechanics, it is impossible to identify these spaces and
to have usual wave function formulation in terms of vectors
in one space. It is onle the special case of the 3--point set on
which all possible topologies are studied that makes it possible to
reduce the construction to one space, since the bra- and
ket--spaces are isomorhic only when $n=3$.

Noncommutativity of some of these matrices
can lead to complementarity (as in the case of Stern-Gerlach
experiment) and to quantum jumps for topologimeter.
Nevertheless one must stress that the example of topologimeter
for the lattice of topologies for three points is the example
of totally new system, different from both classical and
quantum systems. From this one comes to conclusion that
quantum topology can not be thought of as some usual
quantum system described by the wave functions as
vectors in one Hilbert space, but a new formalism
for which a new interpretation is needed.

\section*{Acknowledgments}

One of the authors (A.A.G.) is thankful to CNPq of Brasil for
financial support when completing this work.

\section*{References}
\begin{itemize}
\item[] Birkhoff, G, and J. von Neumann, (1936),
The logic of quantum mechanics,
{\it Annals of Mathematics}, {\bf 37}, 923
\item[] Finkelstein, D., (1963),
{\it Transactions of the New York Academy of Science},
{\bf 25}, 621
\item[] Grib, A.A., and R.R.Zapatrin, (1992),
Topology   Lattice   As Quantum Logic,
{\it International Journal of Theoretical Physics\/},
{\bf 31}, 1093
\item[] Larson R.F., and S.Andima, (1975),
The lattice of topologies: a survey,
{\it Rocky Mountains Journal of Mathematics\/},
{\bf 5}, 177
\item[] Leinaas, J., and R. Myrheim, (1991),
Quantum theories for identical particles,
{\it International Journal of Modern Physics\/},
{\bf B5}, 2573
\item[] von Neumann, J., (1955),
Mathematical Foundations of Quantum Mechanics,
Princeton, New Jersey
\item[] Sorkin, R., (1991),
Finitary Substitutes For Continuous Topology,
{\it International Journal of Theoretical Physics\/},
{\bf 30}, 930
\item[] Zapatrin, R.R., (1993),
Pre-Regge Calculus: Topology Via Logic,
{\it International Journal of Theoretical Physics\/},
{\bf 32}, 779
\item[] Zapatrin, R.R., (1994),
Quantum Logic Without  Negation,
{\it Helvetica  Physica  Acta\/},
{\bf 67}, 188
\item[] D'Espagnat, B., (1976),
{\it Conceptual foundations of quantum mechanics\/},
Benjamin , New York
\end{itemize}
\end{document}